\documentclass[aps,twocolumn,superscriptaddress,showpacs,nofootinbib]{revtex4-2}
\usepackage{graphicx}
\usepackage{dcolumn}
\usepackage{bm}
\usepackage{hyperref}
\usepackage{upgreek}
\usepackage{color}
\usepackage{soul}
\usepackage{epstopdf}
\usepackage{units}
\usepackage{longtable}
\usepackage{floatrow}
\usepackage{verbatim} 
\usepackage{latexsym}
\usepackage{textcomp, gensymb}
\usepackage{floatrow}
\usepackage{amssymb}
\usepackage{ulem}

\makeatletter
\newcommand{\fmarki}{*}
\newcommand{\fmarkii}{*}
\newcommand{\fmarkiii}{\ensuremath{\ddagger}}
\newcommand{\fmarkiv}{\ensuremath{\mathsection}}
\newcommand{\fmarkv}{\ensuremath{\mathparagraph}}
\newcommand{\fmarkvi}{\ensuremath{\|}}
\newcommand{\fmarkvii}{**}
\newcommand{\fmarkviii}{\ensuremath{\dagger\dagger}}
\newcommand{\fmarkix}{\ensuremath{\ddagger\ddagger}}
                
\def\@fnsymbol#1{{\ifcase#1\or \fmarki\or \fmarkii\or \fmarkiii\or \fmarkiv\or \fmarkv\or \fmarkvi\or \fmarkvii\or \fmarkviii\or \fmarkix \else\@ctrerr\fi}}
\makeatother

\begin{document}
\title{Nonlinear planar magnetotransport due to tilted Dirac cones in topological materials}

\author{Arya Thenapparambil}
\let\thefootnote\relax\footnotetext{Corresponding authors:}
\email{arya.thenapparambil@physik.uni-wuerzburg.de}
\affiliation{Faculty for Physics and Astronomy (EP3),
Universität Würzburg, Am Hubland, D-97074 Würzburg, Germany}
\affiliation{Max Planck Institute for Chemical Physics of Solids, Nöthnitzer Straße 40, D-01187 Dresden, Germany}
\affiliation{Institute for Topological Insulators, Am Hubland, D-97074 Würzburg, Germany}

\author{Graciely Elias dos Santos}
\affiliation{Faculty for Physics and Astronomy (EP3),
Universität Würzburg, Am Hubland, D-97074 Würzburg, Germany}
\affiliation{Institute for Topological Insulators, Am Hubland, D-97074 Würzburg, Germany}

\author{Chang-An Li}
\affiliation{Faculty for Physics and Astronomy (TP4),
Universität Würzburg, Am Hubland, D-97074 Würzburg, Germany}

\author{Mohamed Abdelghany}
\affiliation{Faculty for Physics and Astronomy (EP3),
Universität Würzburg, Am Hubland, D-97074 Würzburg, Germany}
\affiliation{Institute for Topological Insulators, Am Hubland, D-97074 Würzburg, Germany}

\author{Wouter Beugeling}
\affiliation{Faculty for Physics and Astronomy (EP3),
Universität Würzburg, Am Hubland, D-97074 Würzburg, Germany}
\affiliation{Institute for Topological Insulators, Am Hubland, D-97074 Würzburg, Germany}

\author{Hartmut Buhmann}
\affiliation{Faculty for Physics and Astronomy (EP3),
Universität Würzburg, Am Hubland, D-97074 Würzburg, Germany}
\affiliation{Institute for Topological Insulators, Am Hubland, D-97074 Würzburg, Germany}

\author{Charles Gould}
\affiliation{Faculty for Physics and Astronomy (EP3),
Universität Würzburg, Am Hubland, D-97074 Würzburg, Germany}
\affiliation{Institute for Topological Insulators, Am Hubland, D-97074 Würzburg, Germany}

\author{Song-Bo Zhang}
\affiliation{Department of Physics, University of Zürich, Winterthurerstrasse 190, CH-8057 Zürich, Switzerland}

\author{Björn Trauzettel}
\affiliation{Faculty for Physics and Astronomy (TP4),
Universität Würzburg, Am Hubland, D-97074 Würzburg, Germany}

\author{Laurens W. Molenkamp}
\email{molenkamp@physik.uni-wuerzburg.de}
\affiliation{Faculty for Physics and Astronomy (EP3),
Universität Würzburg, Am Hubland, D-97074 Würzburg, Germany}
\affiliation{Max Planck Institute for Chemical Physics of Solids, Nöthnitzer Straße 40, D-01187 Dresden, Germany}
\affiliation{Institute for Topological Insulators, Am Hubland, D-97074 Würzburg, Germany}

\begin{abstract}
Nonlinear planar magnetotransport is ubiquitous in topological HgTe structures, both in tensile (topological insulator) or compressively strained layers (Weyl semimetal phase). We show that the common reason for the nonlinear planar magnetotransport is the presence of tilted Dirac cones combined with the formation of charge puddles. The origin of the tilted Dirac cones is the mix of the Zeeman term due to the in-plane magnetic field and quadratic contributions to the dispersion relation. We develop a network model that mimics transport of tilted Dirac fermions in the landscape of charge puddles. The model captures the essential features of the experimental data. It should be relevant for nonlinear planar magnetotransport in a variety of topological and small band gap materials.


\end{abstract}

\maketitle
 The surface states of three-dimensional topological insulators (3DTI) show strong in-plane 
 magnetotransport effects, exhibiting both a strong anisotropic magnetoresistance\cite{CLin,Wang2012,ASulaev,DMahler} as well as its transverse complement\cite{JPJan}, the planar Hall effect\cite{SWiedmann,AATaskin}. A band-structure related origin for these effects was suggested only recently. Zheng et al.\cite{Zheng} argue that the anisotropic backscattering in the surface state necessary for the effects can arise from a tilt of the Dirac cone-dispersion of these states caused by non-linear terms in the band dispersion in combination with the in-plane magnetic field.  

While this model yields a satisfactory understanding of the main experimental results in that 
it is an intrinsic model and does not need to invoke any ad-hoc magnetic impurities\cite{AATaskin},
it does not explain an additional feature of the experimental data, $\it{i.e.}$, the observation of (irregular) fluctuations of the effects' amplitude as a function of the magnitude of the applied magnetic field\cite{DMahler,ASulaev}. An understanding of these fluctuations is important since their presence complicates a convincing demonstration of non-linear Hall and magnetoresistance effects due to Berry curvature dipole\cite{ISodemann,HLegg} in simple double frequency lock-in measurements.
Building upon the idea of Dirac cone tilting\cite{Zheng}, Zhang et al.\cite{SBZhang} recently studied the lateral conductance between two regions of Dirac surface states, separated by a tunnel barrier, in the presence of an in-plane magnetic field and including non-linear terms in the dispersion.  At specific magnetic fields, their model yields a super-resonant transport regime, in which many propagating modes can traverse the barrier potential without backscattering. Evidently, this super-resonant transport can result in pronounced fluctuations of the surface conductance upon tuning the strength and direction of the in-plane field. 

In this paper, we show how the concept of Ref. \cite{SBZhang} can be extended
to satisfactorily describe experimentally observed in-plane magnetotransport fluctuations in HgTe-based devices. We have chosen HgTe-based structures for our transport experiments, since they are nominally undoped and contain virtually no free carriers in the bulk of the material when the Fermi level is adjusted into the surface states, because of their screening properties\cite{CBrune2}.  This allows us to directly study the intrinsic properties of the topological surface states. However, we stress that our findings should hold for any narrow gap system with Dirac-like dispersion. In the HgTe system, we utilize substrate-induced strain to shape the band structure of the epilayer. Because of the inverted band ordering of $\Gamma$\textsubscript{6} and $\Gamma$\textsubscript{8} bands, unstrained bulk HgTe is a topological semimetal, where the semimetallicity stems from the two touching $\Gamma$\textsubscript{8} subbands at the $\Gamma$ point of the Brillouin zone.  
Tensile\cite{CBrune,CBrune2} or compressive\cite{DMahler} strain  lifts this degeneracy by opening a bulk band gap or forming linear crossing points in the band structure, respectively. Thus, by adjusting the strain in the layers through a proper choice of substrate, we can modify the HgTe bandstructure to that of a 3DTI  or of a Weyl semimetal. Note that in both cases, the topological surface state is always present - and dominates the transport - at Fermi energies a few meV away from the $\Gamma$\textsubscript{6} band edge region\cite{DMahler}. Our in-plane magnetotransport experiments establish the remarkably ubiquitous nature of the magneto-fluctuations from surface conduction in the anisotropic magnetoresistance and planar Hall signals. 

To explain our observations, we invoke the presence of sample inhomogeneity due to puddle formation in narrow gap and topological materials. We develop a phenomenological puddle network model. Using the formalism of Ref. \cite{SBZhang}, we show that lateral tunneling of charge carriers between the puddles in parallel magnetic fields does result in prominent fluctuations of the conductance as a function of field amplitude, in good agreement with our experimental observations.

\begin{table}[!htbp]
\caption{\label{tab:table1}%
Type of band structure and fabrication details of all samples.
}
\begin{ruledtabular}
\begin{tabular}{lcccr}
\textrm{Device}&
\textrm{Size ($\mu$m×$\mu$m)}&
\textrm{Type}&
\textrm{Strain}&
\textrm{Gate dielectric}\\
\colrule
A & 600×200 Hall bar & Weyl & 0.26\% & SiO\textsubscript{x}/Si\textsubscript{3}N\textsubscript{4}\\
B & 30×10 Hall bar & Weyl & 0.26\% & HfO\textsubscript{2}\\
C & 600×200 Hall bar & 3DTI & -0.3\% & SiO\textsubscript{x}/Si\textsubscript{3}N\textsubscript{4}\\
D & 5×3 H-bar & Weyl & 0.28\% & SiO\textsubscript{x}/Si\textsubscript{3}N\textsubscript{4}\\
E & 30×10 Hall bar & Weyl & 0.26\% & HfO\textsubscript{2}\\
\end{tabular}
\end{ruledtabular}
\end{table}

Table 1 specifies the details of five 3D HgTe devices used in our in-plane magnetotransport experiments. The samples are grown by molecular beam epitaxy on either CdTe substrates or CdTe/ZnTe superlattices on GaAs as virtual substrates\cite{PLeubner}, allowing us to produce, respectively, tensile (3DTI band structure)  or compressively (Weyl band structure) strained HgTe films of thicknesses 70 nm-120 nm. The corresponding epitaxial strain ranges from -0.3\% (tensile) to 0.28\% (compressive). Devices A,B, and E have been fabricated from the same wafer, devices C an D from two other wafers. Devices with Hall bar and H-bar geometries\cite{CBrune3} were patterned using our established e-beam and optical lithography techniques. Each sample is fitted with a gate insulator layer, covered by a top gate electrode. For samples A, C, and D, a 110 nm-thick SiO\textsubscript{2}/Si\textsubscript{3}N\textsubscript{4} insulator layer grown by plasma-enhanced chemical vapor deposition is used as the dielectric insulator whereas for samples B and E we use a 14 nm-thick HfO\textsubscript{2} insulator layer grown by atomic layer deposition at room temperature. In both cases, the dielectric is covered by a 5 nm Ti/100 nm Au layer. Ohmic contacts to the patterned HgTe mesas are formed by depositing 50 nm/50 nm AuGe/Au layers. All measurements are done at 4.2 K under the application of in-plane magnetic fields of up to 7 T in \textsuperscript{4}He cryostats equipped for low-frequency AC measurements. The longitudinal and Hall resistances of the devices are measured in a four-point configuration using lock-in techniques unless mentioned otherwise.

\begin{figure}[!htbp]
\includegraphics[width=\columnwidth]{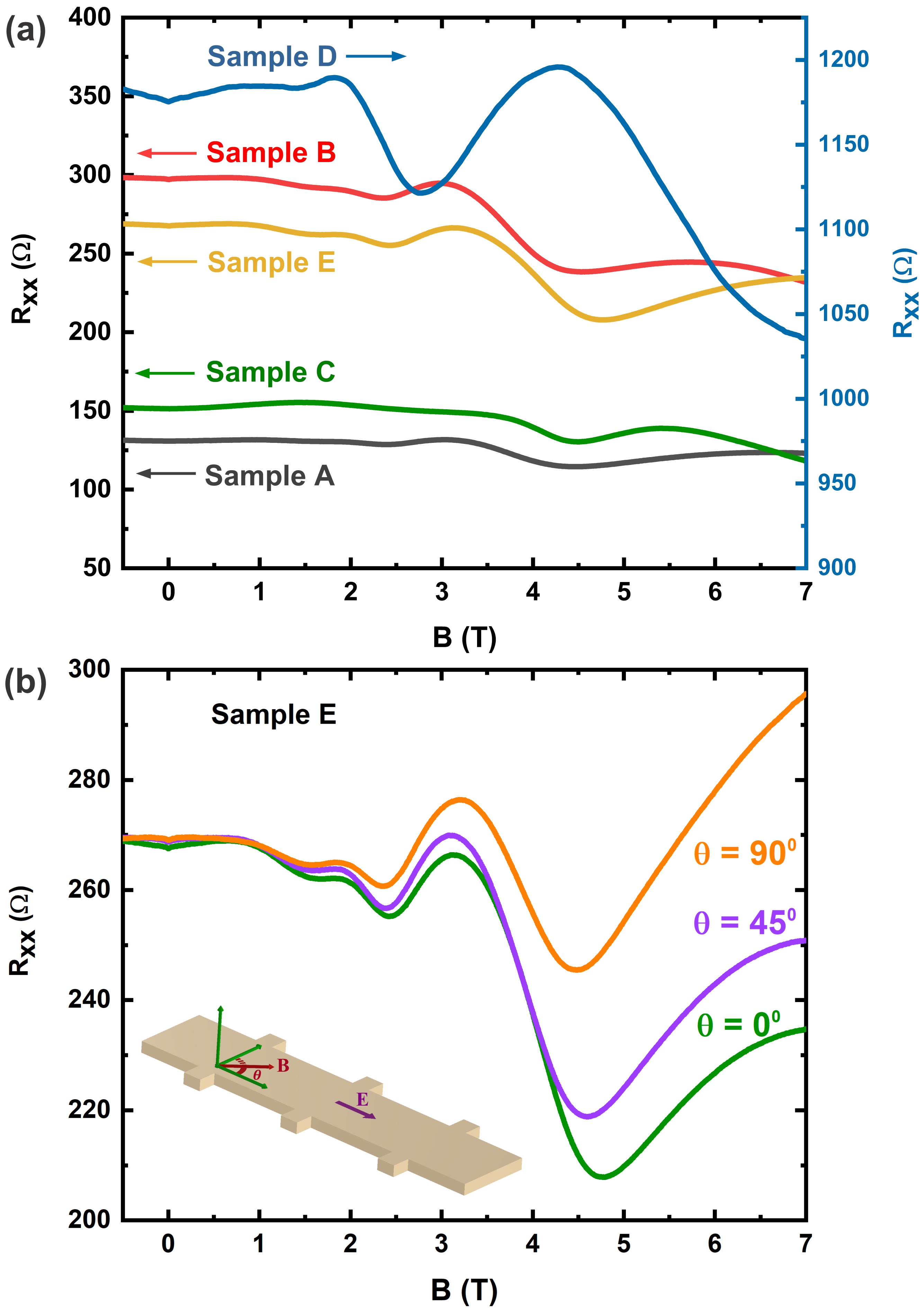}
\caption{\label{} (a) In-plane magnetoresistance of five different samples at an approximate total carrier density of 1 × 10\textsuperscript{12} cm\textsuperscript{-2} as extracted from the low-field Hall data. Sample D (right hand scale) is measured in a quasi four-point configuration and contains a contribution from contact resistance. The left hand scale refers to all other samples. (b) Magnetoresistance of Sample E for $\theta$ = 0$^{\circ}$, 45$^{\circ}$, and 90$^{\circ}$.  }
\end{figure}

The magnetoresistance $\it{R\textsubscript{xx}}$ of all five samples A, B, C, D, and E, for an in-plane magnetic field aligned parallel to the electric field, is shown in Fig. 1(a). Because of the lattice strain, samples A, B, D, and E have a Weyl semimetal bandstructure\cite{JRuan,DMahler} while sample C is a 3DTI\cite{CBrune}. By tuning the gate voltage, all samples are adjusted to a similar total electron density of ca. 1 × 10\textsuperscript{12} cm\textsuperscript{-2}, distributed between top and bottom surface states. At these densities, the transport is totally carried by the topological surface states, as identified by the well-formed quantum Hall plateaus and Shubnikov-de Haas oscillations in out-of-plane field measurements\cite{DMahler,CBrune2, CBrune} (see  also the Supporting Information).  Fig. 1(a) distinctly shows the presence of non-monotonic features in all five longitudinal in-plane magnetoresistance traces. The data obtained from samples A, B, and E (fabricated from the same wafer) show similar fluctuation patterns (which, however, are distinct in the exact field positions). These patterns are more notably different from the fluctuations in the magnetoresistance of samples C and D, fabricated from different wafers. The fluctuating behavior is similar for all inspected carrier densities. Thus the nature of the overall fluctuation pattern apparently varies distinctly between different wafers, and varies only slightly over a single wafer. Also, the position of the fluctuation extrema does not show any periodicity in inverse magnetic fields [see Supporting Information], implying that they are not directly related to bulk Landau level formation. In Fig. 1(b) we plot the magnetoresistance of sample E for three different in-plane angles $\theta$ between the electric and magnetic fields. While the amplitude and position of the fluctuation extrema depend slightly on the angle $\theta$, the overall pattern is very much the same.

\begin{figure}[!htbp]
\includegraphics[width=\columnwidth]{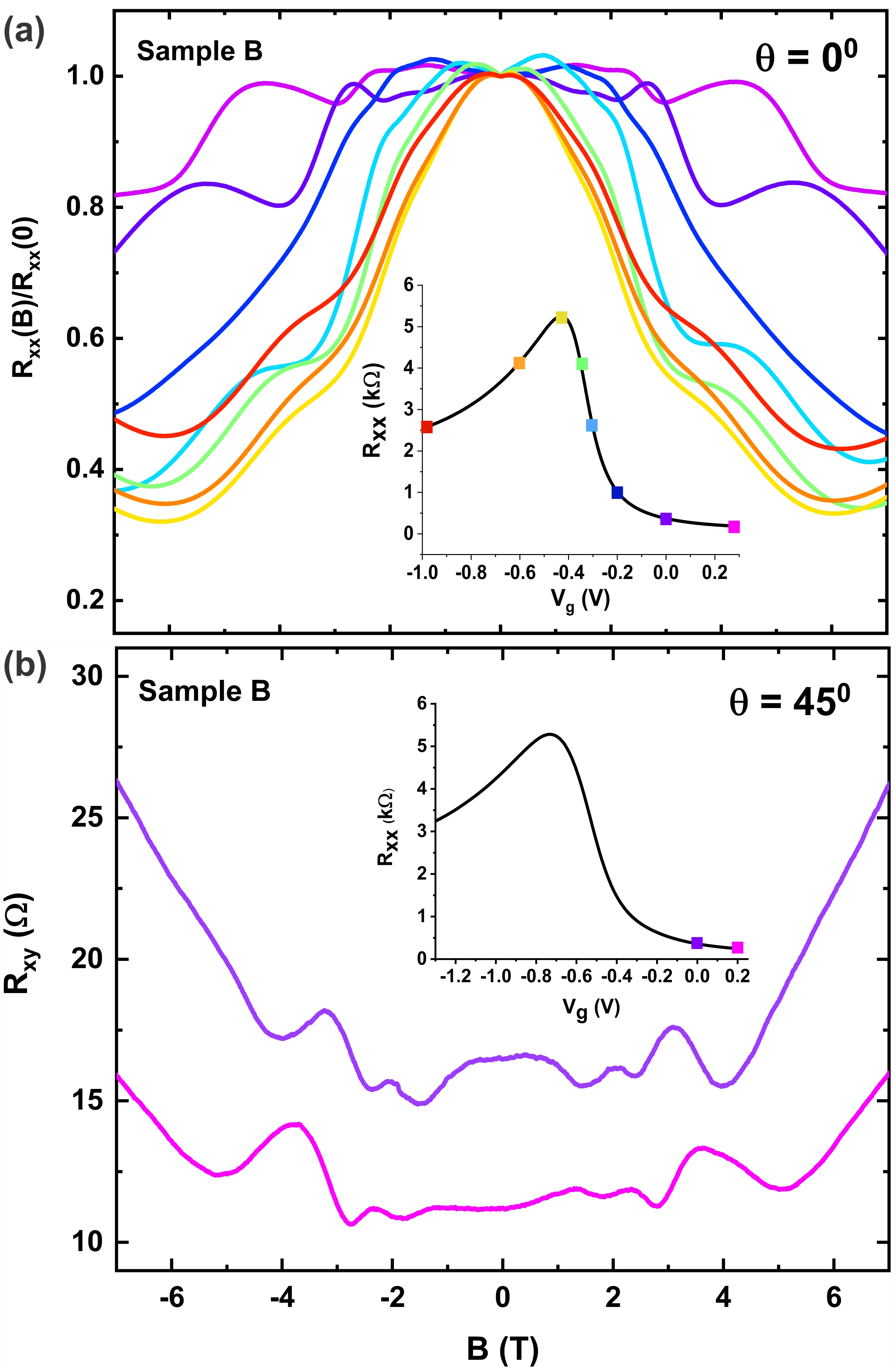}
\caption{\label {} In-plane magnetotransport of a HgTe Weyl semimetal sample. (a) $\it{R\textsubscript{xx}}$ of the first device on Sample B normalized to the value at zero magnetic fields at different $\it{V\textsubscript{g}}$ when $\theta$ = 0$^{\circ}$. (b) Extracted (see main text) $\it{R\textsubscript{xy}}$ of the second device on Sample B at different $\it{V\textsubscript{g}}$, when $\theta$ = 45$^{\circ}$, originating from the planar Hall effect in the topological surface states. The insets to both (a) and (b) show the dependence of $\it{R\textsubscript{xx}}$ as a function of $\it{V\textsubscript{g}}$ for respective Hall bars at zero magnetic fields. }
\end{figure}

 Fig. 2(a) shows the gate voltage dependence of the fluctuations in the in-plane magnetoresistance of a single Hall bar device (sample B, Weyl band structure),
 fabricated along the [110] direction of the epilayer,
 again in parallel magnetic field ($\theta$ = 0$^{\circ}$). The gate voltages are now chosen in a different regime, so that the Fermi level
 is scanned across the Weyl points in the band structure\cite{DMahler}. The inset of Fig. 2(a)
 shows the zero field resistivity of the device. As in Ref. \cite{DMahler}, we infer that the
 Weyl points are located at the maximum of this resistance curve. In this figure,
 $\it{R\textsubscript{xx}}$ is normalized to its zero-field value to facilitate comparison of the behavior of the oscillations with varying gate voltages. The negative magnetoresistance - due to the chiral anomaly\cite{DMahler} in the Weyl semimetal - becomes more pronounced as the Fermi level is tuned from the surface state-dominated n-conducting regime to the Weyl points in the bulk and then diminishes again on the p-conducting side. We find that the positions of the fluctuation extrema tend to shift towards lower in-plane magnetic fields with decreasing total carrier density. Simultaneously, as the Fermi energy is lowered, we also note a reduction in the relative fluctuation amplitudes, implying that the fluctuations emerge and become more prominent as the topological surface states are further populated. Accordingly, we see that the fluctuation amplitude is smallest when the Fermi energy nears the Weyl points in the bulk followed by the p-conducting regime. 
 
 Additionally, in Fig. 2(b), we plot the behavior of the fluctuations in the transverse resistance $\it{R\textsubscript{xy}}$ as a function of field strength, measured on a second Hall bar device from the same wafer, when tuned to the surface state regime. The orientation of the Hall bar in this device is along the [100] crystallographic direction and makes an angle $\theta$ = 45$^{\circ}$ with the applied magnetic field - a configuration where the planar Hall effect originating from an in-plane magnetic field induced anisotropy in the resistance tensor of topological surface states is maximal\cite{JPJan,Zheng}. A linear component in $\it{R\textsubscript{xy}}$(B)  has been subtracted from the raw data of Fig. 2(b) to eliminate the additional contribution from conventional Hall resistance induced by a small out-of-plane field component, caused due to an unavoidable slight misalignment of the sample [see Supporting Information for raw data]. In addition to the non-monotonic behavior, we also observe that the $\it{R\textsubscript{xy}}$ is non-zero at zero fields, which points to a small admixture of $\it{R\textsubscript{xx}}$ in $\it{R\textsubscript{xy}}$, likely due to disorder in the material (cf. our puddle model, below). 
 
 The experimental results presented in Figs. 1 and 2 
 thus demonstrate the ubiquitous and random nature of fluctuations in the in-plane magnetic field response of topological surface state conduction. While these figures focus on Weyl semimetal HgTe samples, in Fig. 3 we look in more detail for similar behavior in the anisotropic magnetoresistance and planar Hall effect of topological surface states in a 3DTI HgTe sample.

\begin{figure}[!htbp]
\includegraphics[width=\columnwidth]{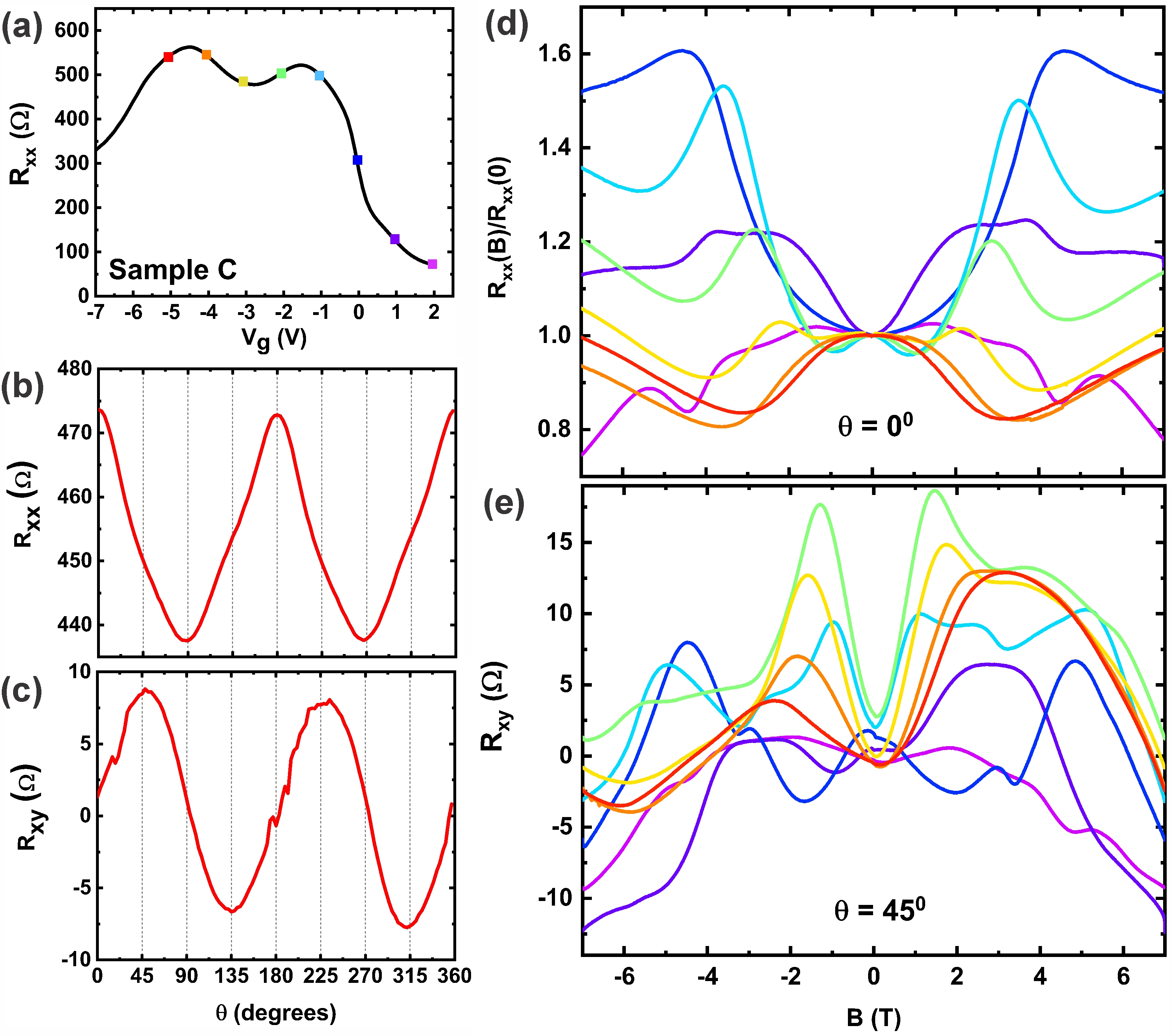}
\caption{\label {}  In plane anisotropic magnetoresistance and planar Hall effect in 3DTI HgTe sample. (a) Gate voltage sweep of $\it{R\textsubscript{xx}}$ of Sample C at zero fields. (b) Longitudinal anisotropic magnetoresistance and (c)Planar Hall effect at varying angles $\theta$, B = 1 T, $\it{V\textsubscript{g}}$= -2 V. (d) $\it{R\textsubscript{xx}}$ of Sample C normalized to the value at zero magnetic fields versus applied in-plane magnetic field when $\theta$ = 0$^{\circ}$ at different $\it{V\textsubscript{g}}$. (e) $\it{R\textsubscript{xy}}$ of Sample C versus applied in-plane magnetic field when $\theta$ = 45$^{\circ}$ at different $\it{V\textsubscript{g}}$. A linear Hall component in the raw data of measured $\it{R\textsubscript{xy}}$ has been subtracted similar to that done in Figure 2 (b). }
\end{figure}

Fig. 3(a) shows the zero-field gate voltage dependence of $\it{R\textsubscript{xx}}$ for Sample C, a Hall bar fabricated from a HgTe 3DTI epilayer. The plot is typical of a high-quality tensile strained HgTe 3DTI sample. The double peak behavior results from the interplay of the high-mobility n-conducting topological surface states and gate-field induced p-conducting massive Volkov-Pankratov surface states\cite{DMahler2}. Subsequently, the magnetotransport of the sample was investigated in a vector magnet applying an in-plane magnetic field of 1 T, yielding the angular dependence of $\it{R\textsubscript{xx}}$ and $\it{R\textsubscript{xy}}$. As shown in Figs. 3(b) and (c), these resistances follow the overall cos$2\theta$ and sin$2\theta$ angular dependence expected\cite{JPJan}  for anisotropic magnetoresistance and planar Hall effect, respectively, yielding a period of 180$^{\circ}$. The amplitude of the effects is of the same order of magnitude as predicted using the model of Ref. \cite{Zheng}.
A full comparison will need to involve a detailed band structure calculation, including proper in-plane g-factors.
At the same time, the traces in Figs. 3(b) and (c) exhibit slight deviations from the expected angular dependencies. This is more obvious in  Figs. 3(d) and (e), which display a noticeable non-monotonic behavior of the in-plane magnetoresistance and planar Hall responses as a function of magnetic field strength, a behavior very similar to our observations in Figs. 1 and 2. This observation once more confirms that the non-periodic fluctuations  in the in-plane magnetic field response arise from the topological surface states. Finally, also Fig. 3(e) displays a finite planar Hall signal at zero in-plane field, again pointing to disorder in the material.

To address the physical origin of the fluctuations with field amplitude in magnetotransport, we invoke the presence of density inhomogeneities in our structures. Previous transport and surface topography studies indicate that the occurrence of such inhomogeneities, resulting in the formation of charge puddles in otherwise undoped, narrow-gap systems like HgTe are inevitable,\cite{Roth2009,LLunczer,MCDartiailh} as they are formed during thin film growth on a substrate with inherent defects. [We note that the origin of these density inhomogeneities in HgTe is rather different from that discussed for Bi-based topological materials,\cite{BSkinner,ARosch} where charge puddles that stem from the high doping density of the materials are taken to be the main source of density inhomogeneity. In contrast to the situation in those compounds, the bandgap of HgTe is not renormalized by the presence of the puddles.] 
Visualization by scanning gate and microwave impedance microscopy in HgTe yield an estimate of size and distribution of the puddles in our material\cite{MKonig,EYMa}. These studies report the presence of charge puddles with a diameter of a few 100 nm and a typical separation of 1 $\mu$m. Our various transport devices thus contain multiple charge puddles.  A schematic, but representative, example of the resulting potential landscape in our devices is shown in Figs. 4(a) and 4(b).

\begin{figure}[!htbp]
\includegraphics[width=\columnwidth]{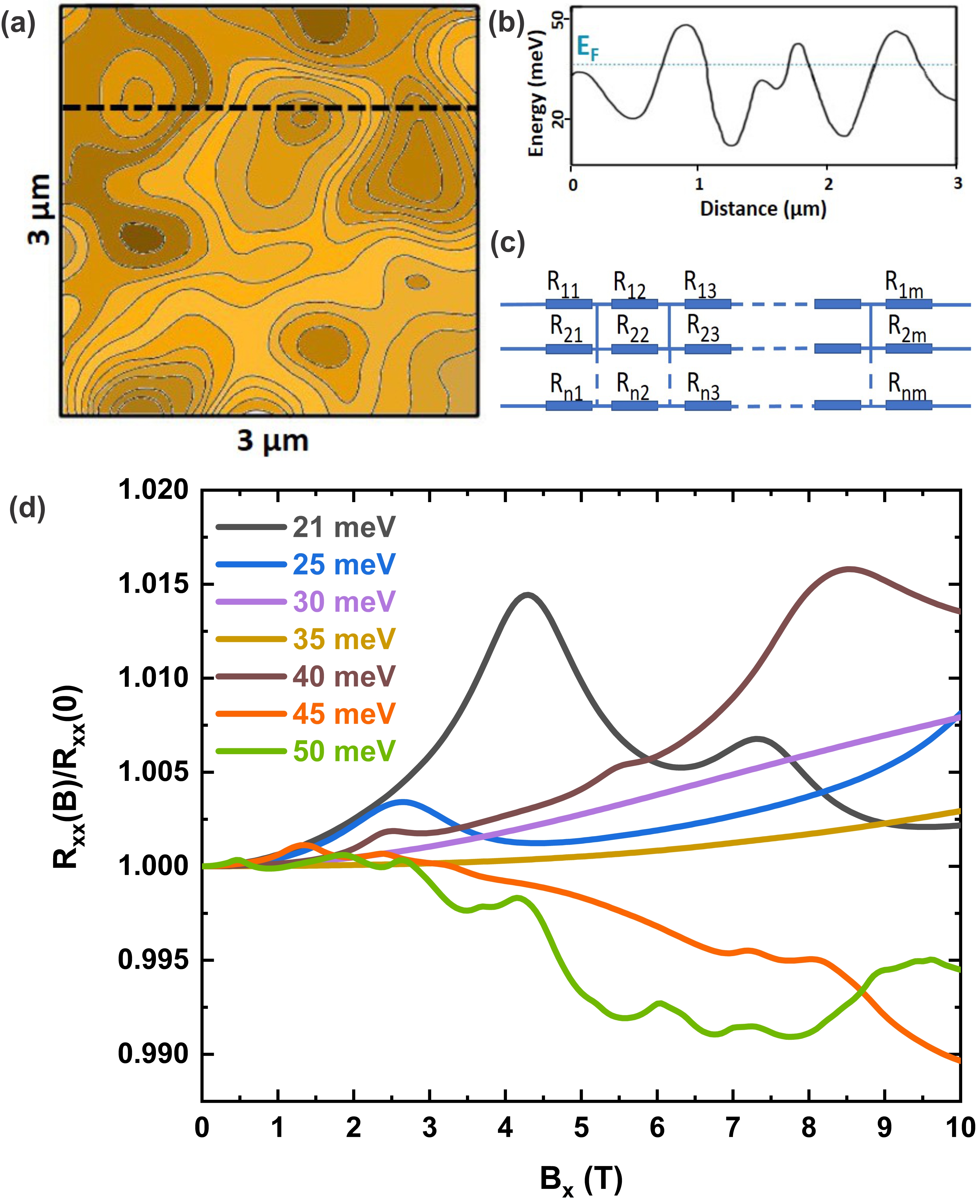}
\caption{\label {}(a) Schematic contour/colour representation of the potential landscape in our samples. Deeper potential inhomogeneities are represented as lighter regions. (b) Cut of the potential landscape along the black dashed line in (a). (c) Resistor network representing the coupled puddles in devices. (d) Resistance fluctuations as a function of magnetic field strength $B$ for different Fermi levels $E_{f}$. For each puddle, we choose the Fermi velocity $v_{f} = 3.9 \times 10^5 $ m.s$^{-1}$, the g-factor $g = 20$, and the curvature parameter $\gamma$ = -0.064 eV.nm$^{2}$ for the Dirac surface states. For the $12 \times 12$ puddle network, the potential barrier of each puddle is assumed to vary in the window $[-40,-20]$ meV, and each puddle size is chosen as $380 \times 380$  $\mathrm{nm^{2}}$. See Supporting Information for a precise definition of the model and its parameters. }
\end{figure}

To model the effect of the puddles on the transport properties of our devices
we use a network model {[}see Fig. 4(c){]} where the low density regions are
implemented as  a chain of independent potential barriers, each barrier being represented as a two-terminal resistor(Details of the model can be found in the Supporting Information). 
The fundamental idea of our simulations
is that each puddle by itself exhibits pronounced fluctuations
in the magneto-conductance due to lateral super-resonant tunneling to its neighbours\cite{SBZhang}.
The network of puddles is then responsible for the full picture of
magnetotransport. The two important parameters relevant for each
puddle are the height of the potential barrier and the size of the puddle. 
We assume that each puddle has a random potential barrier to its neighbor,  drawn from a uniformly distributed range. The transmission probability across the tunnel barrier as a function of in-plane magnetic field is calculated 
using the model of Ref. \cite{SBZhang}. We assume the carriers in the puddles to have the Dirac-type dispersion of the topological surface states,
where the presence of quadratic corrections to the dispersion relation in combination with an in-plane magnetic field tilts the Dirac cones, leading to super-resonant tunneling.

Among the tunnel barriers in each column of the network [see Fig. 4(c)], there will be a dominant one, i.e., the one with the least resistance. The transport of electrons across different adjacent dominant barriers forms a \lq\lq{trajectory}\rq\rq from left to right. The random values of the electrostatic potentials characterizing the dominant puddles lead to different fluctuation patterns (in terms of amplitudes and phases). The \lq\lq{trajectory}\rq\rq then collects a combination of different fluctuation patterns. 

In Fig. 4(d), we show a typical example of the results of the model calculations for a reasonable choice of parameters. In this calculation, we consider a $12 \times 12$ puddle  network, consistent with typical sample sizes in our experiment. Other ranges of parameters are analyzed in the Supporting Information. The results are qualitatively similar to the ones plotted here.
We find that the network model successfully captures the main features of the experimentally observed magnetotransport. In particular, it predicts pronounced fluctuations with field amplitude, comparable in size ($\sim$ 1.5\%) and in field range with the experimental data.

Additionally, the random tunneling between charge puddles in our sample creates inhomogeneity in the conductance distribution sufficient to produce local variations in the electric field direction at any point in the sample. Consequently, in our sample configuration, we always pick up a small fraction of the $\it{R\textsubscript{xx}}$ component while measuring $\it{R\textsubscript{xy}}$, hence, the non-zero value for $\it{R\textsubscript{xy}}$ at zero fields in Figs. 2(b) and 3(e). The simulations that yield a reasonable amplitude of the $\it{R\textsubscript{xx}}$ signal indeed yield up to 10\% pick-up signal in $\it{R\textsubscript{xy}}$.

In conclusion, we have shown that in-plane magnetotransport of topological surface states can be understood by considering Dirac cone tilting in the topological surface states - the tilting itself explains\cite{Zheng} the in-plane magnetic anisotropy and the planar Hall effect, and lateral tunneling between charge puddles\cite{SBZhang} causes an additional fluctuating response on top of these effects. Both effects should
not be limited to topological materials; they should also occur very generally in thin layer materials with Dirac dispersion and sizable spin-orbit coupling, such as, e.g., the transition metal dichalcogenides.
In all these systems the (unavoidable) presence of density fluctuations may seriously complicate the observation of other non-linear magnetotransport phenomena that have been recently proposed.

\begin{acknowledgments}
We thank A. A. Aravindnath and V. L. Müller for assistance in the experiments and L. Fürst and L. Lunczer for growing the HgTe epilayers.
This work was supported by projects SFB 1170 (Project ID 258499086), and through the Würzburg-Dresden Cluster of Excellence on Complexity and Topology in Quantum Matter (EXC 2147, Project ID 390858490); by the Free State of Bavaria through the Institute for Topological Insulators and StMWi project DB001905; and through a Max Planck fellowship at the Max Planck Institute for Chemical Physics of Solids, Dresden.

\end{acknowledgments}

\section*{References}

\bibliography{References}

\end{document}


\title{{Supporting Information\\
Nonlinear planar magnetotransport due to tilted Dirac cones in topological materials}}

\author{Arya Thenapparambil}
\affiliation{Faculty for Physics and Astronomy (EP3),
Universit\"at W\"urzburg, Am Hubland, D-97074 W\"urzburg, Germany}
\affiliation{Max Planck Institute for Chemical Physics of Solids, Nöthnitzer Straße 40, D-01187 Dresden, Germany}
\affiliation{Institute for Topological Insulators, Am Hubland, D-97074 W\"urzburg, Germany}

\author{Graciely Elias dos Santos}
\affiliation{Faculty for Physics and Astronomy (EP3),
Universit\"at W\"urzburg, Am Hubland, D-97074 W\"urzburg, Germany}
\affiliation{Institute for Topological Insulators, Am Hubland, D-97074 W\"urzburg, Germany}

\author{Chang-An Li}
\affiliation{Faculty for Physics and Astronomy (TP4),
Universit\"at W\"urzburg, Am Hubland, D-97074 W\"urzburg, Germany}



\author{Mohamed Abdelghany}
\affiliation{Faculty for Physics and Astronomy (EP3),
Universit\"at W\"urzburg, Am Hubland, D-97074 W\"urzburg, Germany}
\affiliation{Institute for Topological Insulators, Am Hubland, D-97074 W\"urzburg, Germany}

\author{Wouter Beugeling}
\affiliation{Faculty for Physics and Astronomy (EP3),
Universit\"at W\"urzburg, Am Hubland, D-97074 W\"urzburg, Germany}
\affiliation{Institute for Topological Insulators, Am Hubland, D-97074 W\"urzburg, Germany}

\author{Hartmut Buhmann}
\affiliation{Faculty for Physics and Astronomy (EP3),
Universit\"at W\"urzburg, Am Hubland, D-97074 W\"urzburg, Germany}
\affiliation{Institute for Topological Insulators, Am Hubland, D-97074 W\"urzburg, Germany}

\author{Charles Gould}
\affiliation{Faculty for Physics and Astronomy (EP3),
Universit\"at W\"urzburg, Am Hubland, D-97074 W\"urzburg, Germany}
\affiliation{Institute for Topological Insulators, Am Hubland, D-97074 W\"urzburg, Germany}

\author{Song-Bo Zhang}
\affiliation{Department of Physics, University of Zürich, Winterthurerstrasse 190, CH-8057 Zürich, Switzerland}

\author{Björn Trauzettel}
\affiliation{Faculty for Physics and Astronomy (TP4),
Universit\"at W\"urzburg, Am Hubland, D-97074 W\"urzburg, Germany}

\author{Laurens W. Molenkamp}
\affiliation{Faculty for Physics and Astronomy (EP3),
Universit\"at W\"urzburg, Am Hubland, D-97074 W\"urzburg, Germany}
\affiliation{Max Planck Institute for Chemical Physics of Solids, Nöthnitzer Straße 40, D-01187 Dresden, Germany}
\affiliation{Institute for Topological Insulators, Am Hubland, D-97074 W\"urzburg, Germany}
 
 \date{\today}

\setcounter{figure}{0}
\renewcommand{\figurename}{Fig.}
\renewcommand{\thefigure}{S\arabic{figure}}

\maketitle


\section{Additional Experimental Data}

 We use standard Hall measurements to identify the top-gate induced change in carrier density of the sample and the corresponding transport regime when the Fermi level is tuned across the bands. We observe clear quantum Hall plateaus in $\it{R\textsubscript{xy}}$  and Shubnikov-de-Haas oscillations in $\it{R\textsubscript{xx}}$ in perpendicular field measurements for Sample B (when Fermi energy is tuned to the n-region) and Sample C (when Fermi energy is tuned to the gap), as presented in Figs. S1 and S2, respectively. The quantization indicates that the transport is carried exclusively by the two-dimensional topological surface states at these gate voltages\cite{CBrune2, CBrune}. Electron mobilities ranges from $\mu$ $\approx$ $5\times10\textsuperscript{4}$ cm\textsuperscript{2}/V.s in Sample B (compressively strained HgTe grown on (Cd,Zn)Te superlattice on a GaAs substrate) to $\mu$ $\approx$ $3\times10\textsuperscript{5}$ cm\textsuperscript{2}/V.s in Sample C (tensile strained HgTe grown on a CdTe substrate).

\begin{figure}
\includegraphics[width=0.7\linewidth]{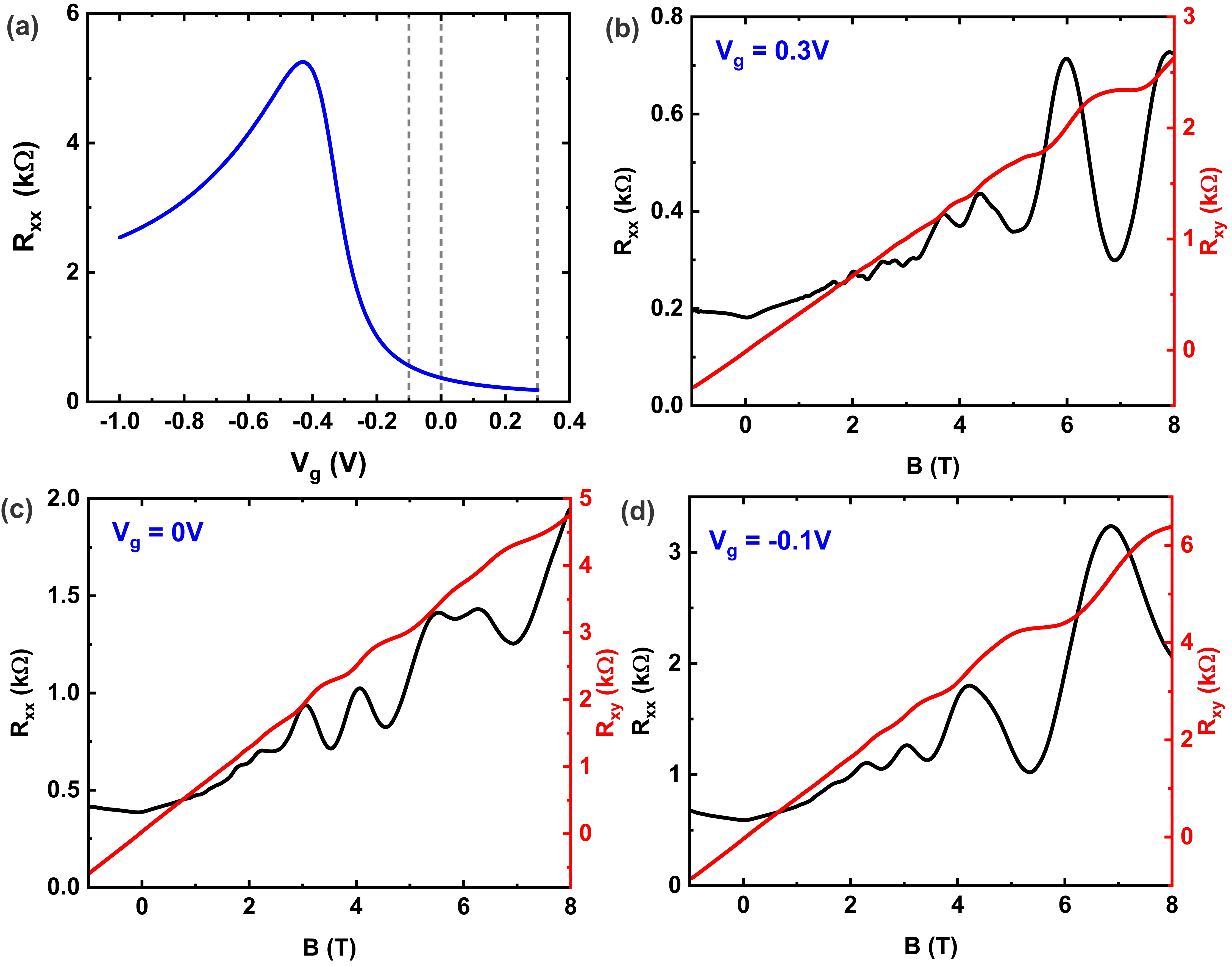}

\caption{Characterization experiments on sample B:  $\it{R\textsubscript{xx}}$ as a function of $\it{V\textsubscript{g}}$ at zero magnetic field. (a)
Magnetotransport (perpendicular field) data for $\it{R\textsubscript{xx}}$ (black) and $\it{R\textsubscript{xy}}$ (red) at 4.2 K as a function of magnetic field at gate voltages (b) $\it{V\textsubscript{g}}$ = 0.3 V (c) $\it{V\textsubscript{g}}$ = 0 V (d)$\it{V\textsubscript{g}}$ = -0.1 V.}
\end{figure}

\begin{figure}
\includegraphics[width=0.7\linewidth]{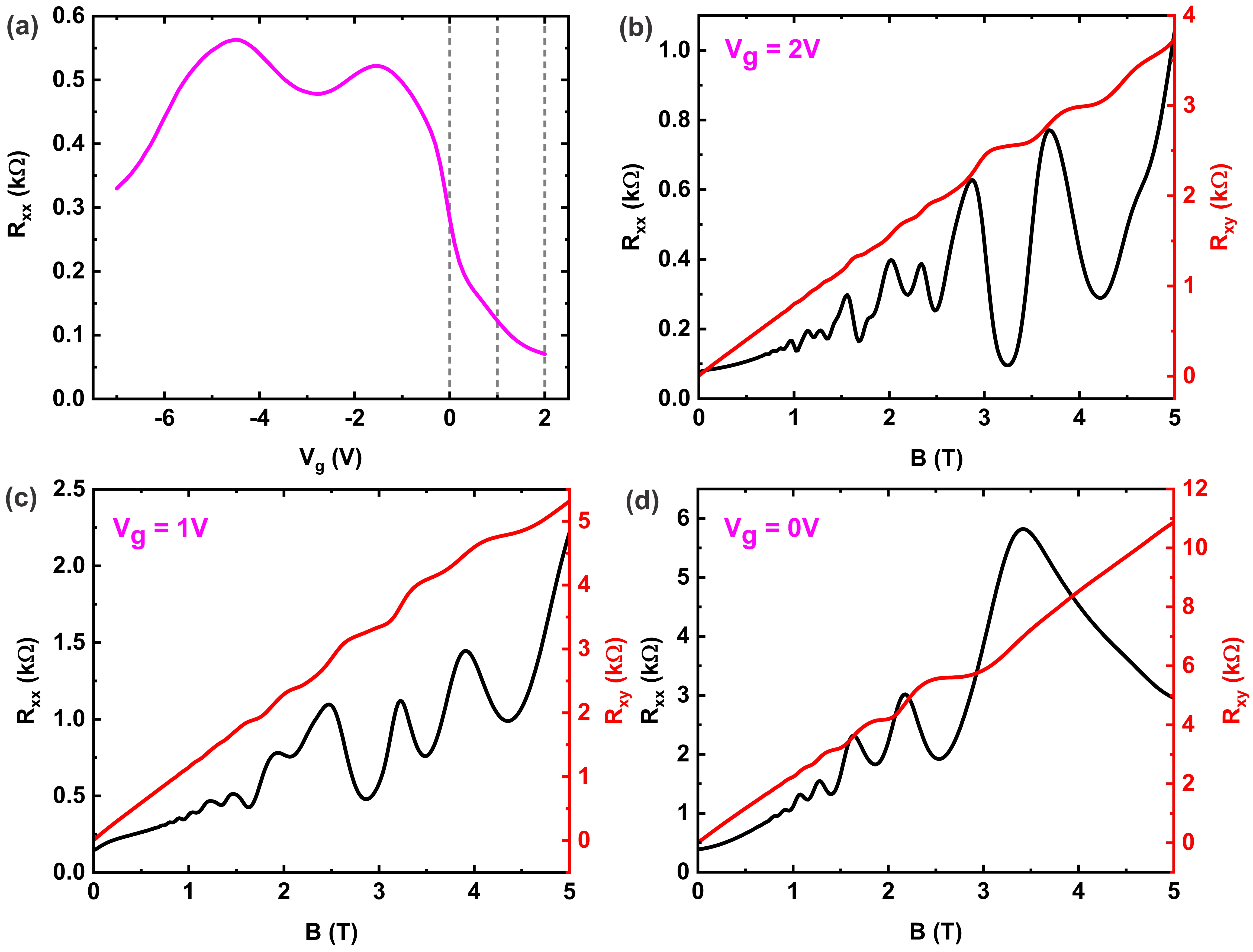}

\caption{Characterization experiments on sample C: $\it{R\textsubscript{xx}}$ as a function of $\it{V\textsubscript{g}}$ at zero magnetic field. (a)
Magnetotransport (perpendicular field) data for  
$\it{R\textsubscript{xx}}$ (black) and $\it{R\textsubscript{xy}}$ (red) at 4.2 K as a function of magnetic field at gate voltages (b)$\it{V\textsubscript{g}}$ = 2 V (c)$\it{V\textsubscript{g}}$ = 1 V (d)$\it{V\textsubscript{g}}$ = 0 V.}
\end{figure}
\FloatBarrier

To demonstrate the absence of periodic behavior in the fluctuating component of the in-plane magnetoresistance, we again plot the $\it{R\textsubscript{xx}}$ of Fig. 1 in the main manuscript, but now  as a function of the inverse field 1/$\it{B}$. The positions of the fluctuation extrema are irregular and non-periodic as can be directly inferred from Fig. S3.   

\begin{figure}[h]
\includegraphics[width=0.7\linewidth]{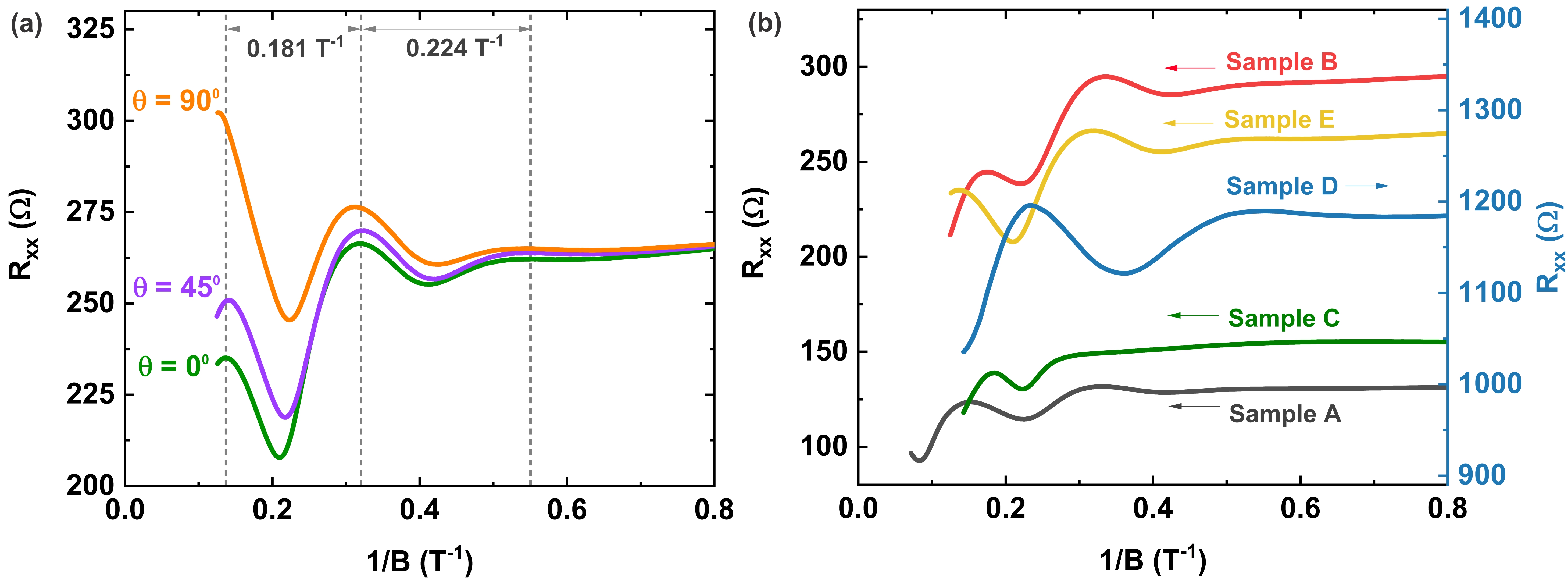}

\caption{(a) $\it{R\textsubscript{xx}}$ of Sample E at different in-plane angles $\theta$ = 0$^{\circ}$, 45$^{\circ}$, and 90$^{\circ}$ shown in Fig. 1(b) of the main text plotted as a function of 1/$\it{B}$. The difference in the separation between the positions of fluctuation maxima as denoted by the  grey dashed lines (for the green curve) is indicative of the lack of any periodicity in the fluctuation extrema in 1/$\it{B}$. (b) $\it{R\textsubscript{xx}}$ of five different samples shown in Fig. 1(a) of the main text plotted as a function of 1/$\it{B}$.}
\end{figure}

In Fig. S4, we present the raw measurement data of the planar Hall signal $\it{R\textsubscript{xy}}$ of Samples B (Fig. S4(a)) and C (Fig. S4(b)). A linear in-$\it{B}$ contribution was subtracted from this raw data to extract the results presented in Figs. 2(b) and 3(e) of the main text. This is done to eliminate any contributions in the signal from the Hall resistance of the device, due to a small perpendicular field component. Sample B was investigated in a measurement stick capable of rotating the sample configuration from in-plane to out-of-plane with respect to the axis of the magnet. In the in-plane field measurements on that sample, the mechanical rotation control was utilized to tweak the orientation of the sample plane with respect to the applied magnetic field in order to minimize the out-of-plane field component. Obviously, there still is a small component remaining, which can be corrected for in a straightforward manner. The situation is worse for sample C, which was investigated in a measurement stick equipped for in-plane rotation only.  This implies an unavoidable slight misalignment of the sample plane when gluing the device to the chip carrier which cannot be corrected mechanically and resulting in a slightly larger contribution of the Hall resistance in the measurements, as evident from Fig. S4(b).   

\begin{figure}
\includegraphics[width=0.7\linewidth]{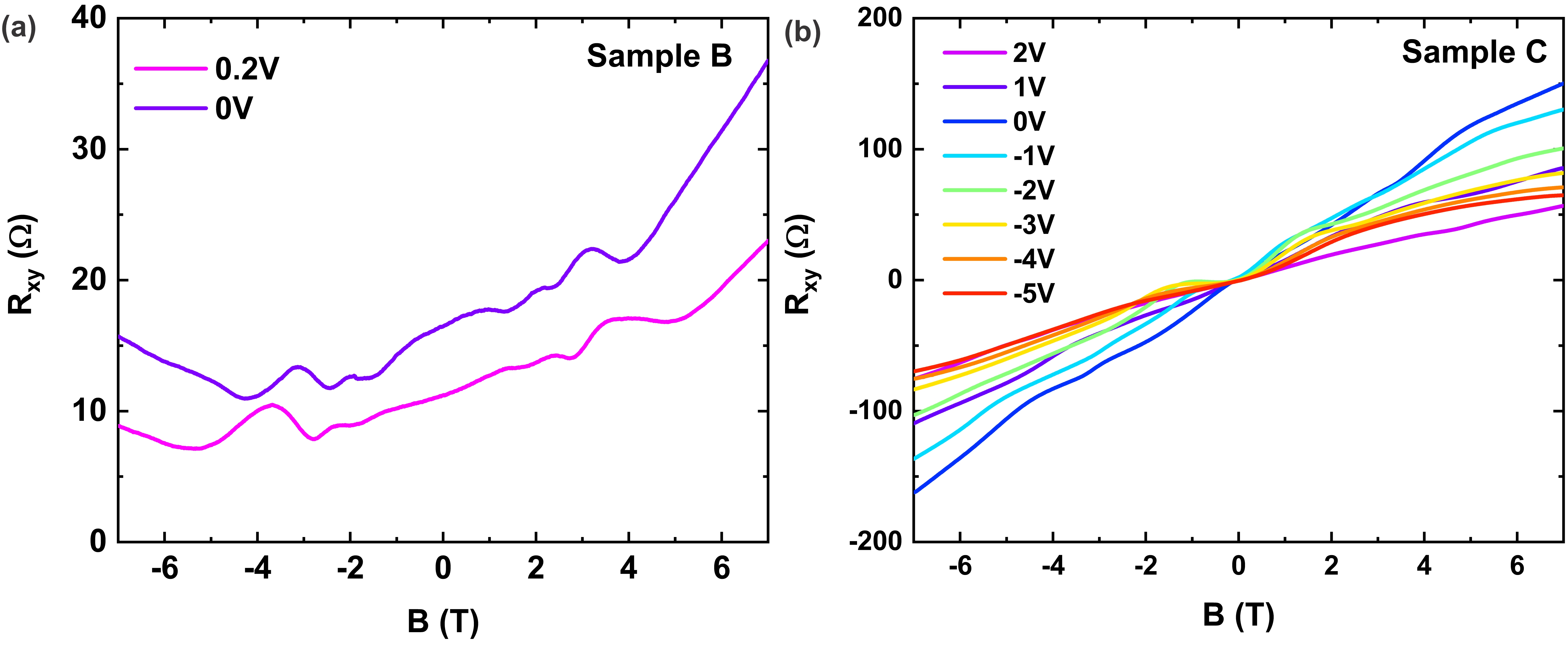}

\caption{Raw data of measured $\it{R\textsubscript{xy}}$ as a function of in-plane magnetic field for $\theta$ = 45$^{\circ}$. (a) $\it{R\textsubscript{xy}}$ of the second device on Sample B  at different $\it{V\textsubscript{g}}$. This is the raw data for Fig. 2(b) (b) $\it{R\textsubscript{xy}}$ of sample C at different $\it{V\textsubscript{g}}$. This is the raw data for Fig. 3(e). A linear component in $\it{R\textsubscript{xy}}$($\it{B}$) has been subtracted from this raw data to obtain Figs. 2(b) and 3(e) presented in the main text.}
\end{figure}

\section{Theory Section of the Supporting Information}

\subsection{Longitudinal magnetotransport}

In this section,\textcolor{blue}{{} }we first recap the calculation
for a single puddle of topological surface states subject to an in-plane
magnetic field published in Ref. \cite{SBZhang}. Then, we show the
simulation results of the network model of puddles. 

\begin{figure}[!h]
\includegraphics[width=0.8\linewidth]{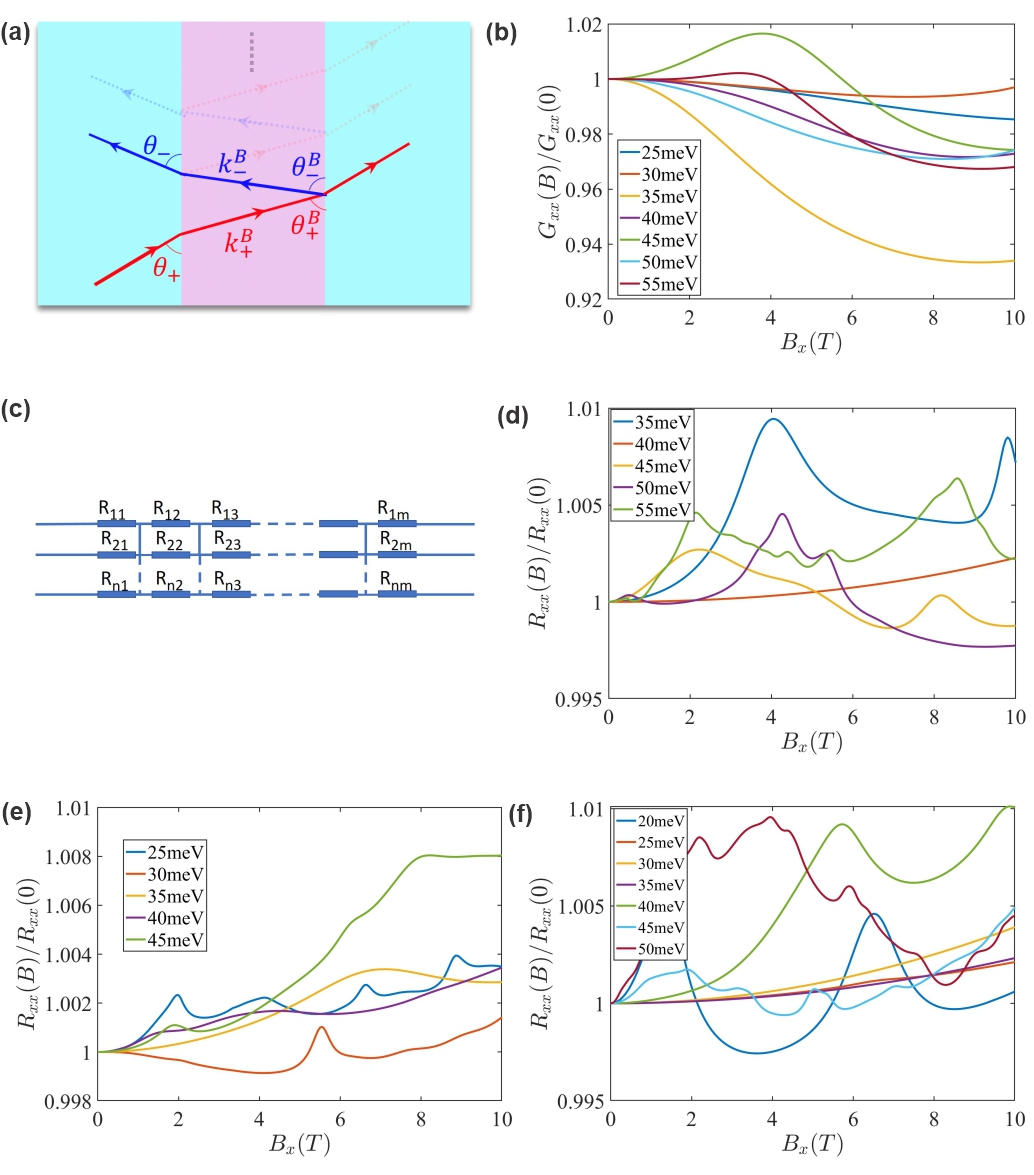}

\caption{(a) Illustration of the scattering process of tilted Dirac surface
states across a puddle (i.e., a barrier junction). (b) Total conductance
$G$ of the network as a function of the in-plane magnetic field strength
$B_{x}$ for different $E_f$. (c) Sketch of the resistor network model. (d) Resistance
$R_{xx}(B)/R_{0}$ of the network model as a function of $B_{x}$ for different $E_f$.
Here, we choose $L_{x}\times L_{y}=380\times380\mathrm{nm}^{2}$,
$N_{x}\times N_{y}=12\times12$, $[U_{t},U_{b}]=[-50,-30]\mathrm{meV}.$
(e) is the same as (d) but for $L_{x}\times L_{y}=380\times500\mathrm{nm}^{2}$
$[U_{t},U_{b}]=[-40,-20]\mathrm{meV}.$ (f) is the same as (d) but
for $N_{x}\times N_{y}=10\times12$. \label{fig:Network}}
\end{figure}

We model a puddle of topological surface states as a potential barrier
with strength $V_{0}$ and length $L$ along $x$-direction, as shown
in the Fig. \ref{fig:Network}(a). It can be described by the following
Hamiltonian

\begin{equation}
H=\mathcal{H}(\hat{{\bf k}})-E_{f}+V({\bf r}),
\end{equation}
where the puddle potential $V({\bf r})$ is $V_{0}$ for $0\leq x\leq L$
and 0 otherwise, $E_{f}$ is the Fermi energy of the surface states.
The effective Hamiltonian for surface states in the presence of an
in-plane magnetic field ${\bf B}=B(\cos\theta,\sin\theta)$ can be
written as\ \cite{SBZhang}
\begin{equation}
\mathcal{H}(\hat{{\bf k}})=\hbar v_{f}(k_{x}s_{y}-k_{y}s_{x})+{\bf t\cdot k},
\end{equation}
where $\hat{{\bf k}}=(k_{x},k_{y})$ is the two-dimensional momentum
near the surface Dirac point, which is a differential operator of
position ${\bf r}=(x,y)$; $v_f$ the Fermi velocity; $s_{x}$ and $s_{y}$
Pauli matrices for spin; ${\bf t}=(t_{x},t_{y})=g\mu_{B}B/2\hbar v_{f}(\gamma\cos\theta-2m\sin\theta,2m\cos\theta-\gamma\sin\theta)$
the tilting vector of the surface Dirac cone due to the concurrence
of the in-plane magnetic field ${\bf B}$ and the quadratic momentum
corrections; $g$ the effective g-factor of surface states; $\mu_{B}$
the Bohr magneton; the effective mass $m$ = -0.108 eV.nm$^{2}$ and the curvature parameter $\gamma$ = -0.064 eV.nm$^{2}$ for surface states in HgTe.

We study the transport properties of the system by employing the scattering
approach\ \cite{SBZhang}. For simplicity, we assume the transverse
momentum $k_{y}$ as a good quantum number and use it to label the
transport modes. A finite confinement in $y$-direction (as considered
in the puddles) results in a quantization of $k_{y}$ determined by
the width $L_{y}$, i.e., $k_{y}=n\pi/L_{y}$ where $n$ being an
integer.\textcolor{black}{{} The transmission probability across the
junction [see Fig. S5(a)] can be found analytically.} The expression for the transmission
probability is too cumbersome to be displayed here [see Ref. \cite{SBZhang}]. However, for large
barriers with $|V_{0}|\gg E_{F}$, it can be simplified to
\begin{equation}
T_{k_{y}}=\frac{1-\cos(\theta_{+}-\theta_{-})}{1-\sin\theta_{+}\sin\theta_{-}-\cos[(k_{+}^{B}-k_{-}^{B})L]\cos\theta_{+}\cos\theta_{-}}.\label{eq:Transmission Prob}
\end{equation}
Here, the angles $\theta_{+}$ ($\theta_{-}$) indicate the direction
of incoming (outgoing) electrons, and $k_{\pm}^{B}$ are the wave
vectors in the puddle region, as sketched in Fig. \ref{fig:Network}(a).
From Eq. (\ref{eq:Transmission Prob}), we see that the potential
barrier becomes transparent for all transport modes (labeled by $k_{y}$)
when the resonance condition $\sin[(k_{+}^{B}-k_{-}^{B})L]=0$ is
satisfied. This result is the super-resonant transport \cite{SBZhang}.
Analogously, we find that all transport modes have instead the lowest
transmission probabilities when the anti-resonance condition $\cos[(k_{+}^{B}-k_{-}^{B})L]=0$
is fulfilled. The conductance $G$ at zero temperature can be evaluated
as the sum of the transmission probability of all modes as

\begin{equation}
G=\frac{e^{2}}{h}\sum_{k_{y}}T_{k_{y}}.
\end{equation}

As shown in Ref. \cite{SBZhang}, super-resonance and super-anti-resonance
conditions can be satisfied alternately as we increase the in-plane
magnetic field $B$. Consequently, we observe pronounced in-plane
magneto-fluctuations, as shown in Fig. \ref{fig:Network}(b). Note
that such pronounced fluctuations persist even when $V_{0}$ becomes
comparable to the Fermi energy $E_{f}$ and for an arbitrary magnetic
field direction $\theta$.

Next, we proceed to discuss the puddle network model that simulates
the observed anomalous magneto-fluctuations. The fabricated samples
are not homogeneous due to the formation of puddles. These puddles
generally have shallow potentials compared with the Fermi energy,
and their spatial sizes fluctuate slightly. Based on these experimental
guidances, we propose a phenomenological model, i.e., a parallel resistor
network model. Explicitly, we consider each puddle as a resistor and
connect them as a network, as shown in Fig. \ref{fig:Network}(c).
As discussed in the main text, the two important parameters relevant
for each puddle are potential barrier and spatial size. Generally,
we therefore have different parameter scenarios to consider: (i) fixed
puddle sizes $L_{x}\times L_{y}$ but fluctuating potential barrier;
(ii) fixed potential barrier potential barrier but fluctuating puddle
sizes $L_{x}\times L_{y}$; (iii) both the potential barrier and puddle
size fluctuate. 

Let us first consider case (i). We assume that each puddle has a random
potential drawn from a uniform distribution range $[U_{b},U_{t}]$,
and its spatial size is fixed at $L_{x}\times L_{y}$. The fundamental
idea of our simulations is that each puddle by itself exhibits pronounced
fluctuations in conductance due to super-resonant tunneling.
Moreover, the network of puddles is responsible for the net fluctuation
patterns. The longitudinal resistance can be directly obtained from
the conductance as $R=1/G$.

Among the puddles in each column of the network, there is typically
a dominant one, i.e., the one with least resistance. Consequently,
the tunneling of electrons between different adjacent dominant puddles
forms a ``trajectory'' from left to right. The total resistance
of the network model is evaluated as
\begin{equation}
R_{xx}(B)=\sum_{m=1}^{N_{x}}\frac{1}{\sum_{n=1}^{N_{y}}\frac{1}{R_{nm}(B)}}.
\end{equation}
As shown in Fig. 3(d) of the main text, the network model gives rise
to pronounced fluctuations in the magneto-conductance that are comparable with the ones
observed in experiments. We normalize the resistance by $R_{0}=R_{xx}(B=0)$
in the plots.

We now provide more simulations by tuning the parameters
in the network model to show that the magneto-fluctuations are a general
and robust phenomenon. In Fig. \ref{fig:Network}(d), we take a different
barrier potential range, i.e., $[U_{t},U_{b}]=[-50,-30]\ \mathrm{meV}$
as compared to Fig. 3(d) in the main text. We find that the magneto
fluctuations persist. In Fig. \ref{fig:Network}(e), we instead change
the spatial size of each puddle to $L_{x}\times L_{y}=380\times500\ \mathrm{nm}^{2}$.
In Fig. \ref{fig:Network}(f), we modify the number of puddles to
$N_{x}\times N_{y}=10\times12$. Hence, the magneto-fluctuations persist
as a general property in different parameter regimes. All of these
cases show irregular but pronounced fluctuations with field amplitude, capturing
the main features of the observed magneto-transport. 

\begin{figure}[!h]
\includegraphics[width=0.8\linewidth]{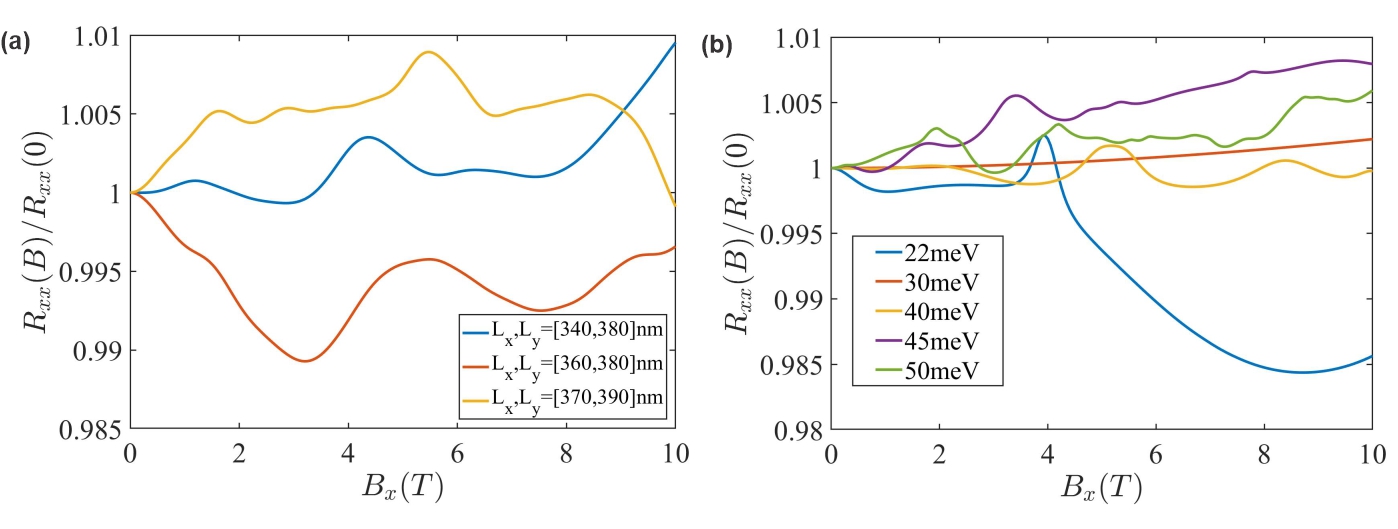}

\caption{Persistence of fluctuations in the magneto-conductance under different choices of relevant
parameters. (a) We choose the size of puddle size $L_{x}$ and $L_{y}$
independently in different distributed ranges. The Fermi energy is
fixed at $E_{f}=45$ meV, and the potential barrier is $V_{0}=30$
meV. (b) The puddle sizes $L_{x}$ and $L_{y}$ are independently
chosen from the range $[340,380]$ nm, and the potential barrier $V_{0}$
of each puddle is chosen from the uniformly distributed range $[-40,-20]$
meV for different values of $E_f$ as specified in the legend. \label{fig:Network-2}}
\end{figure}
\FloatBarrier

Next, we show the persistence of these fluctuations in case (ii)
and case (iii). In Fig. \ref{fig:Network-2}(a), for case (ii), we
fix the puddle potential $V_{0}$ but choose puddles of random size
from a uniformly distributed range. In Fig. \ref{fig:Network-2}(b),
for case (iii), we consider both the potential and size of the puddles
to be random. The corresponding simulating results are qualitatively
similar to the ones in Fig. \ref{fig:Network}. Therefore, the magneto-fluctuations appear in different scenarios by varying the relevant
parameters (such as potential barrier and puddle sizes) in the puddle
network model.

\subsection{Finite Hall resistance at zero fields}

\begin{figure}[h]
\includegraphics[width=0.4\linewidth]{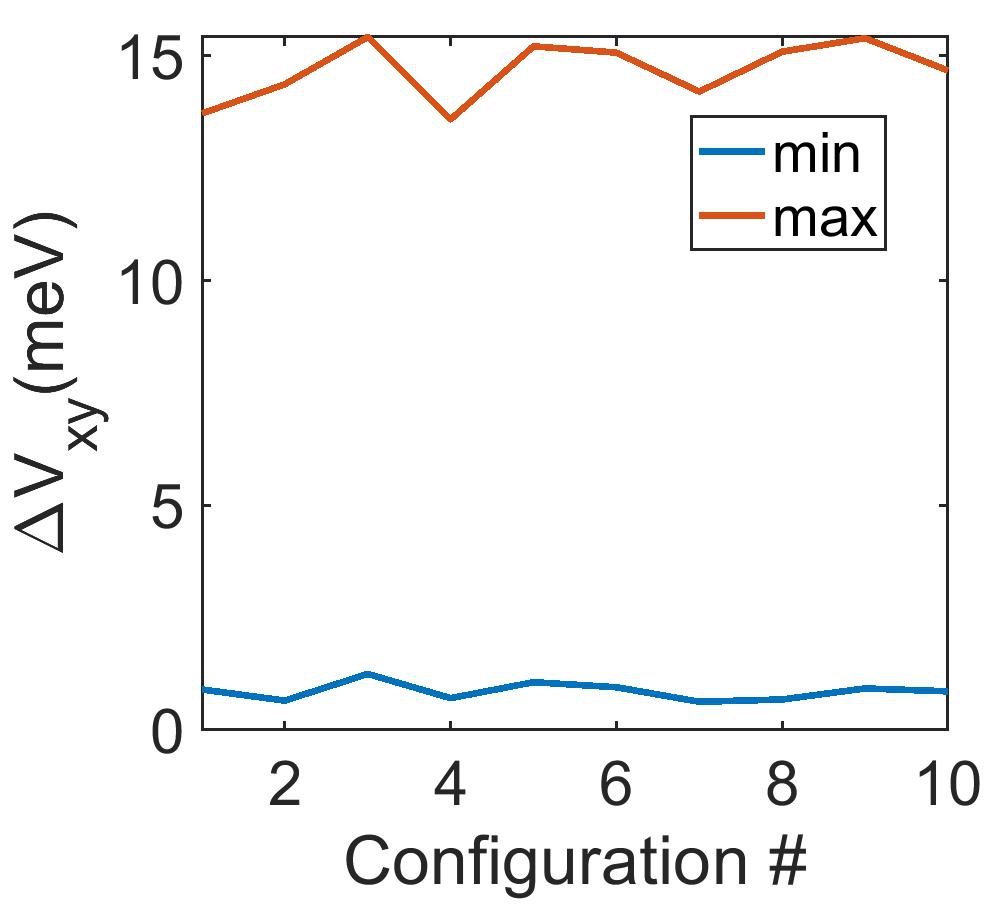}

\caption{Fluctuation of the potential difference $\Delta V_{xy}^{\mathrm{max/min}}$
for different configurations in the network model. We choose $N_{x}\times N_{y}=12\times12$,
$[U_{t},U_{b}]=[-40,-20]\mathrm{meV}.$ The magnetic field strength
is at zero. \label{fig:Network-1}}
\end{figure}

Experimentally, we observe a nonzero Hall resistance $R_{xy}$ at
zero magnetic field $B=0$, as shown in Fig. 2(b) in the main text.
This observation indicates the existence of a finite transversal potential
difference $V_{xy}\neq0$. We provide a possible explanation of it
by the presence of puddles, consistent with the network mode explained
above. Assume that the network model is composed of $N_{x}$ columns
and $N_{y}$ rows of puddles. Each puddle is labeled by a coordinate
$(i\hat{x},j\hat{y})$. The potential of each puddle varies in a range
$[U_{t},U_{b}]$ randomly. Focusing on each column of puddles in the
network model, there exists a nonzero potential difference along the
$\hat{y}$ direction. Let us consider the potential difference between
nearest neighbor puddles, which can be calculated as $\Delta V_{y,ij}=|V_{i,j}-V_{i,j+1}|$
for fixed column index $i$. These values fluctuate for different
configurations. We determine the maximum $\Delta V_{y,i}^{\mathrm{max}}$
and minimum $\Delta V_{y,i}^{\mathrm{min}}$ in each column. Then
we average over different columns in the network model (over different
column index $i$),
\begin{equation}
\Delta V_{xy}^{\mathrm{max/min}}=\frac{1}{N_{x}}\sum_{i}V_{y,i}^{\mathrm{max/min}}.\ 
\end{equation}
As shown in the Fig. \ref{fig:Network-1}, the transversal potential
difference along $\hat{y}$ direction fluctuates around some fixed
values. It is about $1$ meV for $\Delta V_{xy}^{\mathrm{min}}$ and
$14$ meV for $\Delta V_{xy}^{\mathrm{max}},$ respectively. When
compared with the potential window range, the relative percentage
is about $2.5\%\sim5\%$ for $\Delta V_{xy}^{\mathrm{min}}$ and $35\%\sim70\%$
for $\Delta V_{xy}^{\mathrm{max}}$. These estimates for the transversal
potential fluctuations are in accordance with the experimental observations
of a finite $R_{xy}$ at zero magnetic field.

